\documentclass[10pt,journal,twoside]{IEEEtran}
\usepackage{graphicx}
\usepackage{amsmath}
\usepackage{amsfonts}
\usepackage{amssymb}
\usepackage{psfrag}
\usepackage{cite}
\usepackage{xcolor}
\usepackage{multirow}
\usepackage{subfigure}
\usepackage[bookmarksnumbered]{hyperref}
\usepackage{multicol}
\newtheorem{theorem}{\underline{Theorem}}
\newtheorem{lemma}{\underline{Lemma}}

\newtheorem{assumption}{\underline{Assumption}}
\newtheorem{problem}{\underline{Problem}}

\newtheorem{proposition}{\underline{Proposition}}
\newtheorem{definition}{\underline{Definition}}

\newcommand{\QED}{{\rm $\blacksquare$}}

\begin{document}
\title{Noncoherent Multiuser Massive SIMO for Low-Latency Industrial IoT Communications 
}
\author{Zheng Dong$^*$, He Chen$^*$, Jian-Kang Zhang$^\dag$, and Branka Vucetic$^*$\\	
$^*$The University of Sydney, Australia, emails:\{zheng.dong, he.chen, branka.vucetic\}@sydney.edu.au\\
$^\dag$McMaster University, Canada, email:jkzhang@mail.ece.mcmaster.ca
}
\vspace{-30pt}
\maketitle
\begin{abstract}
In this paper, we consider a multiuser massive single-input multiple-output (SIMO) enabled Industrial Internet of Things (IIoT) communication system. To reduce the latency and overhead caused by channel estimation, we assume that only the large-scale fading coefficients are available. We employ a noncoherent maximum-likelihood (ML) detector at the receiver side which does not need the instantaneous channel state information (CSI). For such a massive SIMO system, we present a new design framework to assure that each transmitted signal matrix can be uniquely determined in the noise-free case and be reliably estimated in noisy cases. The key idea is to utilize a new concept called the uniquely decomposable constellation group (UDCG) based on the practically used quadrature amplitude modulation~(QAM) constellation. To improve the average error performance when the antenna array size is scaled up, we  propose a max-min Kullback-Leibler (KL) distance design by carrying out optimization over the transmitted power and the sub-constellation assignment. Finally, simulation results show that the proposed design outperforms significantly the existing max-min Euclidean distance based method in terms of error performance. Moreover, our proposed approach also has a better error performance than the conventional coherent zero-forcing (ZF) receiver with orthogonal training for cell edge users.
\end{abstract}


\section{Introduction}
The fourth stage of industrial revolution, also termed Industry 4.0 or Industrial Internet of Things (IIoT), is a new vision that in general refers to the introduction of advanced Internet technologies in industrial control and manufacturing, with the aim of significantly boosting the flexibility, versatility, usability and efficiency of future smart factories~\cite{Drath14}. There is a general consensus that providing powerful and pervasive connectivity between machines, people and objects in industrial environments will be essential to realize this new vision. 

Connectivity in factories has until now been dominated by wired systems, which has been preferred for its real-time capability and reliability. Wireless systems are mainly applied in noncritical use cases such as monitoring of conditions, which poses comparatively low communication performance requirements since communication failure will not lead to serious accidents (e.g., economic losses and safety problems)~\cite{Huang18}. The main reason behind this is that stringent performance in terms of relibility and latency required by critical use cases are several orders of magnitude better than what is achievable by today’s wireless technologies~\cite{Chen2018mag}. On the other hand, there are increasing necessity for the use of wireless solutions in critical use cases because it reduces cost for large-scale deployment, enables flexible communication needed in smart production, and opens new fields of application, such as the control of moving objects together with their simultaneous localization and allocation. As such, there is an urgent need to develop ultra-reliable low-latency wireless communications for critical IIoT use cases. Recently, massive multiple-input multiple output (MIMO) technology, which refers to the usage of a large number of antennas in wireless systems to harness the spatial diversity, has been regarded indispensable for providing ultrahigh reliability~\cite{ Popovski18}. In light of this, how to develop low-latency massive MIMO systems has become one of the most important research problems in the field.

Motivated by the above facts, we consider an uplink multiuser massive single-input multiple-output (SIMO) enabled IIoT communication system. 
We note that the channel estimation is a major obstacle to reduce the transmission latency due to the following two reasons: 1) the estimation of channel state information (CSI) is commonly done by transmitting some known orthogonal pilot symbols, the minimum number of which is no less than the number of transmitter antennas~\cite{Gershman06}. This will cause significant delay when the user number is large; 
2) different from most conventional communication systems which commonly have very long data packets, the control/data packets in IIoT applications are typically very short. Thus, the overhead caused by channel estimation is non-negligible and will reduce the information rate significantly~\cite{Popovski16}.  In our design, to avoid the overhead and latency caused by channel estimation, we adopt a noncoherent maximum-likelihood (ML) receiver, where no instantaneous CSI is needed at all the nodes. 
The proposed new transmission framework is based on the concept called uniquely decomposable constellation group (UDCG) with QAM constellations. Simulation results show that our proposed design outperforms the currently available max-min Euclidean distance based method. Our contributions can be summarized as follows:
\begin{itemize}
\item In our design, by jointly considering two time slots, constellations with phase information (e.g., QAM) can be used by all the users~\cite{Varanasi01,Aazhang03tit}, while existing works on single-user or multiuser noncoherent massive SIMO considered only one time slot, the information can only be modulated on the amplitudes of the input signals, resulting in a low spectral efficiency~\cite{Goldsmith16tit, Popovski16tsp, Gao18iotj}.

\item We present an explicit construction of UDCG for \emph{any} number of users. 
Unlike design in~\cite{Chevillat81} which is based on the trellis coding structure, whose received constellation is very complicated and hard to decode, the received constellation in our design has a simple geometric structure.
\end{itemize} 

\section{The System Model and Noncoherent Maximum-Likelihood (ML) Detector}
\subsection{The System Model and Noncoherent ML Detector}\label{sec:systemmodel}
We consider a massive SIMO system consisting of $K$ single-antenna users transmitting simultaneously to the BS with $M$ ($M\gg K$) receiving antennas on the same time-frequency grid. By using a discrete-time complex baseband-equivalent model, the received signal at the antenna array of BS in the $t$-th time slot $\mathbf{y}_t=[y_{1,t},\ldots, y_{M,t}]^T$ can be expressed by
\begin{align*}
\mathbf{y}_t=\mathbf{H}\mathbf{x}_t +{\boldsymbol\xi}_t,
\end{align*}
where $\mathbf{x}_t=[ x_{1,t},\ldots, x_{K,t}]^T$ represents the transmitted signals from all the $K$ users, ${\boldsymbol\xi}_t$ is an additive circularly-symmetric complex Gaussian (CSCG) noise vector with covariance $\sigma^2 \mathbf{I}_M$. We let $\mathbf{H}=\mathbf{G}\mathbf{D}^{1/2}$ denote the $M\times K$ complex channel matrix between the receiver antenna array and all the users, where $\mathbf{G}$ characterizes the small-scale fading caused by local scattering while $\mathbf{D} ={\rm diag}\{\beta_1,\cdots, \beta_K\}$, $\beta_k>0$ captures the propagation loss due to distance and shadowing. All the entries of $\mathbf{G}$ are assumed to be i.i.d. complex Gaussian distributed with zero mean and unit variance. The channel coefficients are assumed to be in block fading which are quasi-static in the current block and change to other independent values in the next block with a channel coherence time $T_c \ge K$. We consider a space-time block modulation (STBM)~\cite{Aazhang03tit} scheme over $T$ time slots and the received signal vectors can be stacked together into a matrix given by
\begin{align}\label{eqn:vecrecsignal}
\mathbf{Y}_T=\mathbf{H}\mathbf{X}_T +{\mathbf\Xi}_T,
\end{align}
where $\mathbf{Y}_T=[\mathbf{y}_1,\ldots, \mathbf{y}_T]$, $\mathbf{X}_T=[\mathbf{x}_1, \ldots, \mathbf{x}_T]$ and ${\mathbf \Xi}_T=[\boldsymbol{\xi}_1,\cdots, \boldsymbol{\xi}_T]$.

\begin{assumption}\label{assumption1}
 Throughout this paper, we adopt the following assumptions:
\begin{enumerate}
\item  The small scale channel fading matrix $\mathbf{G}$ is completely unknown to the BS and all the users, while the large scale fading matrix $\mathbf{D}$ is available to all the nodes;
\item The transmitted signals are subject to an instantaneous average power constraint\footnote{Note that our design can be directly extended to the case with peak power constraint.}: $\mathbb E \{|x_{k,t}|^2\} \le P_k$, $k=1,\ldots, K$, $t=1,\ldots, T$. For convenience, we assume that the users are labeled  such that $P_1\beta_1\le \ldots \le P_K \beta_K$. 
\end{enumerate} 
\end{assumption}

We consider a noncoherent ML detector which is optimal for uniformly distributed discrete input signals in terms of error probability. First, we note that~\eqref{eqn:vecrecsignal} can be reformulated by $\mathbf{Y}_T^H=\mathbf{X}_T^H \mathbf{D}^{1/2}\mathbf{G}^H +{\boldsymbol \Xi}_T^H$. Then, 
 the vectorized version of the received signal can be written as
\begin{align*}
\mathbf{y}={\rm vec}(\mathbf{Y}_T^H)=(\mathbf{I}_{M}\otimes \mathbf{X}_T^H \mathbf{D}^{1/2}){\rm vec}(\mathbf{G}^H) +{\rm vec}(\boldsymbol{\Xi}_T^H).
\end{align*}
As all the entries of $\mathbf{G}$ and $\boldsymbol \Xi$ are i.i.d. CSCG distributed, we immediately have $\mathbb E[\mathbf{y}] =\mathbf{0}$, and the covariance matrix of $\mathbf{y}$ can be calculated by 
$\mathbf{R}_{\mathbf{y}|\mathbf{X}_T}=\mathbb E[\mathbf{y}\mathbf{y}^H]
=\mathbf{I}_M \otimes (\mathbf{X}_T^H \mathbf{D}\mathbf{X}_T +\sigma^2 \mathbf{I}_T)$. 
Then, the conditional distribution of the received signal $\mathbf{y}$ at BS for any transmitted signal matrix $\mathbf{X}_T$ is given by 
\begin{align*}
p({\mathbf{y}|\mathbf{X}_T})=\frac{1}{\pi^{K M}\det({\mathbf{R}_{\mathbf{y}|\mathbf{X}_T}})} \exp\Big(-\mathbf{y}^H \mathbf{R}_{\mathbf{y}|\mathbf{X}_T}^{-1}\mathbf{y}\Big).
\end{align*}
The noncoherent ML detector aims to estimate the transmitted information carrying matrix from the received signal vector $\mathbf{y}$ by solving the optimization problem:
$\widehat{\mathbf{X}}_T ={\arg\max}_{\mathbf{X}_T}~\ln p({\mathbf{y}|\mathbf{X}_T})$, or equivalently,

\begin{align}\label{eqn:MLdetector}
\widehat{\mathbf{X}}_T={\arg\min}_{\mathbf{X}_T}~\mathbf{y}^H \mathbf{R}_{\mathbf{y}|\mathbf{X}_T}^{-1}\mathbf{y}+\log \det(\mathbf{R}_{\mathbf{y}|\mathbf{X}_T}).
\end{align}

From~\eqref{eqn:MLdetector}, we can observe that the detector relies on the sufficient statistic of the transmitted signal matrix: $\mathbf{R}_{\mathbf{y}|\mathbf{X}_T}=\mathbf{I}\otimes (\mathbf{X}_T^H \mathbf{D}\mathbf{X}_T +\sigma^2 \mathbf{I}_T)$. The detailed discussion is given in the following subsection.
\subsection{Unique Identification of the Transmitted Signal Matrix}
In this section, we first identify what conditions the transmitted signal matrix must satisfy to enable the unique identification of the transmitted signal matrix $\mathbf{X}_T$.
We can observe from~\eqref{eqn:MLdetector} that, to achieve reliable communication between all the users and BS in the considered massive SIMO system, the receiver must be able to uniquely determine each transmitted signal matrix ${\mathbf X}_T$ once $\mathbf{R}=\mathbf{X}_T^H \mathbf{D}\mathbf{X}_T$ has been received, which can be formally stated as follows:
\begin{proposition}\label{proposition:UFCM}
	Any reliable communications for the multiuser massive SIMO system described in~\eqref{eqn:vecrecsignal} require that, for the transmitted signal matrix selected from ${\mathcal M}^{K \times T}\subseteq \mathbb C^{K \times T}$, if and only if there exist any two signal matrices $\mathbf{X}_T, \widetilde{\mathbf{X}}_T\in {\mathcal M}^{K \times T}$ satisfying $\mathbf{X}_T^H \mathbf{D}\mathbf{X}_T= \widetilde{\mathbf{X}}_T^H \mathbf{D}\widetilde{\mathbf{X}}_{T}$, then we have ${\mathbf{X}}_T=\widetilde{\mathbf{X}}_T$.\hfill\QED
\end{proposition}
The proof is omitted due to space limitation. Inspired by Proposition~\ref{proposition:UFCM},  to aid our system design, we introduce the following concept called uniquely-factorable multiuser space-time modulation (UF-MUSTM) as follows:
\begin{definition}\label{def:udmustm}
A multiuser space-time modulation codebook $\mathcal{S}^{K\times T} \subseteq \mathbb C^{K \times T}$ is said to form a UF-MUSTM codebook if for any pair of codewords $\mathbf{S}, \widetilde{\mathbf{S}} \in \mathcal{S}^{K\times T}$ satisfying $\mathbf{S}^H \mathbf{S} = \widetilde{\mathbf{S}}^H \widetilde{\mathbf{S}}$, we have $\mathbf{S} =\widetilde{\mathbf{S}}$.	\hfill\QED
\end{definition}

Definition~\ref{def:udmustm} motivates us to design a UF-MUSTM codebook for the considered noncoherent massive SIMO, which will be given in Sec.\,\ref{SecIII}. Therefore, our primary task in the rest of this paper is to propose a new method for the systematic design of such UF-MUSTM ${\mathcal S}^{K\times T}$. 

We note that, most of the existing space-time code designs consider point-to-point MIMO systems, where all the transmitting antennas are connected to the same transmitter. As a result, the transmitted information carrying signals are accessible by all the antennas where unitary space-time code design can be employed~\cite{Varanasi01, Aazhang03tit}. However, in our considered MUSTM noncoherent massive SIMO system, the signals transmitted from different users are not allowed to fully collaborate, which greatly limits the codebook design. Moreover, the performance analysis for non-unitary codeword of MUSTM is very challenging as shown in~\cite{Varanasi01}.


\section{The Multiuser Massive SIMO System with Noncoherent ML Receiver}\label{SecIII}
In this section, we present a UF-MUSTM framework with a slot-by-slot ML detection receiver. We find that when the number of receiver antennas increases, the pairwise error probability (PEP) between the two codewords will be dominated by the Kullback-Leibler (KL) distance between them. Motivated by this fact, a max-min KL distance design is proposed by performing optimization on sub-constellation assignment and power allocation among all the users. 
\subsection{The KL Distance between the Transmitted Space-Time Modulation Codewords}
In practice, the computational complexity of the optimal noncoherent ML detector described in~\eqref{eqn:MLdetector} can be prohibitively high when the size of $\mathbf{X}_T$ is large. 
To reduce the receiver complexity, our main idea is to use a small block size for the ML receiver. If only one time slot is involved in the ML detector in~\eqref{eqn:MLdetector}, i.e., we consider $T=1$, the correlation matrix $\mathbf{R}=\mathbf{X}_T^H \mathbf{D}\mathbf{X}_T$ degenerate into a real scalar $\mathbf{x}_1^H\mathbf{D}\mathbf{x}_1=\sum_{k=1}^K \beta_k |x_{k,1}|^2$, where the phase information of the transmitted symbols is lost and information bits from all the users can only be modulated on the amplitudes of the transmitted symbols. However, such a design typically has a very low spectral efficiency~\cite{Goldsmith16tit, Popovski16tsp, Gao18iotj}. To improve the spectrum efficiency by allowing constellation with phase information be transmitted by all the users, we need at least two time slots~\cite{Varanasi01,Aazhang03tit}. 
Motivated by the above observation, we consider the case with $T=2$, where the transmitted signals from the first and second time slots are represented by $\mathbf{X}_2=[\mathbf{x}_1, \mathbf{x}_2]$. We also denote 
 $\mathbf{R}_{\mathbf{y}|\mathbf{X}_2}=\mathbf{I}_M\otimes \mathbf{R}_2$, in which
\begin{align}\label{eqn:correlation2by2}
&\mathbf{R}_2=\mathbf{X}_2^H \mathbf{D}\mathbf{X}_2 +\sigma^2 \mathbf{I}_2
=\begin{bmatrix}
a & c\\
c^* & b
\end{bmatrix},
\end{align}
where $a=\mathbf{x}_1^H \mathbf{D} \mathbf{x}_1+\sigma^2$, $b=\mathbf{x}_2^H \mathbf{D} \mathbf{x}_2 +\sigma^2$, and $c=\mathbf{x}_1^H \mathbf{D} \mathbf{x}_2$, with $ab>|c|^2$.
By~\eqref{eqn:correlation2by2}, we immediately have
\begin{align}\label{eqn:correlation2by2inv}
&\mathbf{R}_2^{-1}
=\frac{1}{ab-|c|^2} \begin{bmatrix}
b & -c \\
-c^* & a
\end{bmatrix}.
\end{align}
By inserting~\eqref{eqn:correlation2by2}, \eqref{eqn:correlation2by2inv} into～\eqref{eqn:MLdetector}, the ML receiver can be given by:
\begin{align}\label{eqn:simplifiedMLreceiver}
\widehat{\mathbf{X}}_2&={\arg\min}_{\mathbf{X}_2}~\mathbf{y}^H \mathbf{R}_{\mathbf{y}|\mathbf{X}_2}^{-1}\mathbf{y}+\log \det(\mathbf{R}_{\mathbf{y}|\mathbf{X}_2})\nonumber\\
&={\arg\min}_{\mathbf{X}_2}~\frac{ a \|\mathbf{y}_2\|^2 + b \|\mathbf{y}_1\|^2 -2 \Re(c \mathbf{y}_2^H \mathbf{y}_1)}{ab-|c|^2}\nonumber\\
&\qquad \qquad \qquad \qquad\qquad \qquad + M \ln \big(ab-|c|^2\big),
\end{align}
where $\mathbf{y}_1$ and $\mathbf{y}_2$ are the received signal vectors in the first and second time slots, respectively.	
It can be observed that the diagonal entries in~\eqref{eqn:correlation2by2} are $a=\mathbf{x}_1^H\mathbf{D}\mathbf{x}_1=\sum_{k=1}^K \beta_k |x_{k,1}|^2$ and $b=\mathbf{x}_2^H\mathbf{D}\mathbf{x}_2=\sum_{k=1}^K \beta_k |x_{k,2}|^2$, in which the phase information is lost, while the off-diagonal term is $c=\mathbf{x}_1^H \mathbf{D} \mathbf{x}_2 =\sum_{k=1}^K \beta_k x_{k,1}^*x_{k,2}=\sum_{k=1}^K \beta_k |x_{k,1}||x_{k,2}|\exp\big (j\arg(x_{k,2}) -j\arg(x_{k,1})\big )$, suggesting that we can transmit a known reference signal vector $\mathbf{x}_1$ in the first time slot and then transmit the information bearing signal vector $\mathbf{x}_2$ to enable a ``differential-like" transmission. 
The exact PEP is extremely hard to evaluate for the matrix $\mathbf{X}_2$ given above~\cite{Varanasi01}.
Moreover, the exact expression for the pairwise error probability (PEP) do not seem to be tractable for optimization. Inspired by the Chernoff-Stein Lemma, when the number of receiver antennas $M$ goes to infinity, the PEP will goes to zero exponentially where the exponent determined by the KL distance~\cite{Aazhang03tit}. Hence, in this paper, we propose to maximize the minimum KL distance between the conditional distributions of the received signals corresponding to different input signals.

We now calculate the KL distance between the received signals induced by the  transmitted signals matrices $\mathbf{X}_2=[\mathbf{x}_1, \mathbf{x}_2]$ and $\widetilde{\mathbf{X}}_2=[\tilde{\mathbf{x}}_1, \tilde{\mathbf{x}}_2]$, which is also the expectation of the likelihood function between two received signals vectors. In essence, the likelihood function between the received signal vectors corresponding to the two transmitted signals convergence in probability to the KL-distance with the increase of the number of receiver antennas.  More specifically, the KL-distance between the received signal corresponding to transmitted matrix $\mathbf{X}_2$ and $\widetilde{\mathbf{X}}_2$ is given by:
\begin{subequations}
	\begin{align*}
	&\mathcal{D}_{\rm KL}^{(M)}(\mathbf{X}_2 ||\widetilde{\mathbf{X}}_2)=\mathbb E_{f({\mathbf{y}|\mathbf{X}_2})} \bigg[ \ln \Big ( \frac{f({\mathbf{y}|\widetilde{\mathbf{X}}_2})}{f({\mathbf{y}|\mathbf{X}_2})} \Big ) \bigg]\\
	&\!=\!\mathbb E_{f({\mathbf{y}|\mathbf{X}_2})} \bigg[ \ln \Big (\frac{\det({\mathbf{R}_{\mathbf{y}|\mathbf{X}_2}}) }{\det({\mathbf{R}_{\mathbf{y}|\widetilde{\mathbf{X}}_2}})}\Big )\! +\!\Big ( \mathbf{y}^H \mathbf{R}_{\mathbf{y}|\widetilde{\mathbf{X}}_2}^{-1}\mathbf{y} -\mathbf{y}^H \mathbf{R}_{\mathbf{y}|\mathbf{X}_2}^{-1}\mathbf{y}\Big )\bigg]\\
	&\!=\!\mathbb E_{f({\mathbf{y}|\mathbf{X}_2})} \bigg[ {\rm tr}\Big (\big ( \mathbf{R}_{\mathbf{y}|\widetilde{\mathbf{X}}_2}^{-1} - \mathbf{R}_{\mathbf{y}|\mathbf{X}_2}^{-1}\big )\mathbf{y}\mathbf{y}^H \Big )\bigg]+\ln \Big ( \frac{\det({\mathbf{R}_{\mathbf{y}|\mathbf{X}_2}}) }{\det({\mathbf{R}_{\mathbf{y}|\widetilde{\mathbf{X}}_2}})}\Big )\\
	&\!=\! {\rm tr}\Big (\big (\mathbf{R}_{\mathbf{y}|\widetilde{\mathbf{X}}_2}^{-1} - \mathbf{R}_{\mathbf{y}|\mathbf{X}_2}^{-1}\big )\mathbf{R}_{\mathbf{y}|\mathbf{X}_2} \Big )\!+\!\ln \Big (\frac{\det({\mathbf{R}_{\mathbf{y}|\mathbf{X}_2}}) }{\det({\mathbf{R}_{\mathbf{y}|\widetilde{\mathbf{X}}_2}})}\Big )\\
	&=M\, \mathcal{D}_{\rm KL}(\mathbf{X}_2 ||\widetilde{\mathbf{X}}_2),
	\end{align*}
\end{subequations}
in which
\begin{align}
&\mathcal{D}_{\rm KL}(\mathbf{X}_2 ||\widetilde{\mathbf{X}}_2)= {\rm tr} \big[ (\mathbf{X}_2^H \mathbf{D}\mathbf{X}_2+\sigma^2 \mathbf{I}_2)(\widetilde{\mathbf{X}}_2^H \mathbf{D}\widetilde{\mathbf{X}}_2 +\sigma^2 \mathbf{I}_2)^{-1} \big] \nonumber\\
& - \ln\Big[ \det\big ((\mathbf{X}_2^H \mathbf{D}\mathbf{X}_2+\sigma^2 \mathbf{I}_2)(\widetilde{\mathbf{X}}_2^H \mathbf{D}\widetilde{\mathbf{X}}_2 +\sigma^2 \mathbf{I}_2)^{-1}\big )\Big]-2.\label{eqn:KLdistanece}
\end{align}
We can observe that $\mathcal{D}_{\rm KL}(\mathbf{X}_2 ||\widetilde{\mathbf{X}}_2)$ is actually the KL-distance if there is only one receiver antenna. Due to the assumption of the independence of channel coefficients, and the KL distance with $M$ antennas $\mathcal{D}_{\rm KL}^{(M)}(\mathbf{X}_2 ||\widetilde{\mathbf{X}}_2)$ is $M$ times of $\mathcal{D}_{\rm KL}(\mathbf{X}_2 ||\widetilde{\mathbf{X}}_2)$.

\subsection{Multiuser Space-Time Modulation witn QAM Division}
The main objective of this subsection is to propose a new QAM division based MUSTM design framework for the considered massive SIMO system. This design is based on the uniquely decomposable constellation group (UDCG) originally proposed in~\cite{dong16jstsp}, building upon the commonly used spectrally efficient QAM signaling.
Now, we introduce the definition of UDCG as follows:
\begin{definition}\label{def:audcg}
	A group of constellations $\{\mathcal{X}_k\}_{k=1}^K$ form a UDCG, denoted by $\big\{ \sum_{k=1}^K x_k: x_k \in \mathcal{X}_k\big\}=\uplus_{k=1}^K \mathcal{X}_k = \mathcal{X}_1 \uplus \ldots \uplus \mathcal{X}_K$, if there exist two groups of $x_k, \tilde x_k \in \mathcal{X}_k$ for $k=1, \cdots, K$ such that $\sum_{k=1}^K x_k =\sum_{k=1}^K \tilde x_k$, then we have $x_k =\tilde x_k$ for $k=1, \cdots, K$.~\hfill\QED
\end{definition}

As QAM constellation is commonly used in modern digital communications, which has a simple geometric structure, we now give the following construction of UDCG based on QAM constellation. For simplicity, we consider that each user is using the 4-QAM constellation~\footnote{The case with general QAM constellation is much more complicated and will be left as a future work.}.

\begin{lemma}\label{lemma:UDCG} \emph{The UDCG with multilevel 4-QAM constellations}: The $4^K$-ary square QAM constellation $\mathcal{Q} = \big\{[\pm (m-\frac{1}{2})\pm j(n-\frac{1}{2}) ]d : m,n =1,\ldots, 2^{K -1}\big\}$, with $d$ being the minimum Euclidean distance between the constellation points, can be uniquely decomposed into the sum of $K$ multilevel 4-QAM sub-constellations $\{{\mathcal X}_k\}_{k=1}^K$ denoted by $\mathcal{Q} = \uplus_{k=1}^K \mathcal{X}_k$, where
$\mathcal{X}_k =\big\{(\pm \frac{1}{2}\pm \frac{j}{2})\times 2^{k-1 }d\big\}$  for $k =1, \ldots,K$.
\hfill\QED
\end{lemma}

With the help of UDCG, we are now ready to propose a QAM division based UF-MUSTM for the considered massive SIMO system with a noncoherent ML receiver given in~\eqref{eqn:simplifiedMLreceiver}. The structure of each transmitted signal matrix is given by $\mathbf{X}_{2}=[\mathbf{x}_1, \mathbf{x}_2]=\mathbf{D}^{-1/2}\mathbf{\Pi} {\mathbf S}_2$, in which
\begin{align}\label{eqn:sigmtx}
\mathbf{S}_2&= [\mathbf{s}_1, \mathbf{s}_2]= \begin{bmatrix}
	\frac{1}{\sqrt{p_1}}& \sqrt{p_1} s_{1}\\
	\frac{1}{\sqrt{p_2}}& \sqrt{p_2} s_{2}\\
	\vdots & \vdots\\
	\frac{1}{\sqrt{p_K}} & \sqrt{p_K} s_{K}
	\end{bmatrix}.
\end{align}
In our design, the diagonal matrix $\mathbf{D}^{-1/2}$ is used to compensate for the large scale fading between different users. The vector $\mathbf{p}=[p_1, \ldots, p_K]$ is introduced to adjust the relative transmitting power between all the users and $\mathbf{s}=[s_1, \ldots, s_K]$ is the information carrying vector. We let  $s_k\in \mathcal{X}_k$ where $\mathcal{X}_k$ constitute a UDCG with sum-QAM constellation $\mathcal{Q}$ such that $\mathcal{Q} =\uplus_{k=1}^K \mathcal{X}_k$ as defined in Lemma~\ref{lemma:UDCG} and $\mathbb E[|s_k|^2]=E_k d^2$, with $E_k=2^{2k-3}$, $k=1,\ldots, K$. 
The matrix $\mathbf{\Pi}=[\mathbf{e}_{\pi{(1)}}, \ldots, \mathbf{e}_{\pi(K)}]^T$ is a permutation matrix, where  $\mathbf{e}_k$ denotes a standard basis column vector of length $K$ with 1 in the $k$-th position and 0 in other positions. $\pi:\{1,\ldots,K\} \to \{1,\ldots, K\}$ is a permutation over $K$ elements characterized by
$\begin{pmatrix}1 &2 &\ldots & K\\\pi(1) &\pi(2)&\ldots&\pi(K)\end{pmatrix}$. We also let  $\pi^{-1}:\{1,\ldots,K\} \to \{1,\ldots, K\}$ be a permutation such that $\pi^{-1}(\pi(k))=k$ for $k=1,\ldots, K$. From the above definition, we immediately have $\mathbf{\Pi}^T\mathbf{\Pi}=\mathbf{I}_K$.

For transmitted signal matrices $\mathbf{X}_{2}$, we have the following desired properties:
\begin{proposition}\label{prop:udcg}
Consider $\mathbf{X}_{2}=\mathbf{D}^{-1/2}{\mathbf \Pi}{\mathbf S}_2$ and $\widetilde{\mathbf{X}}_{2}=\mathbf{D}^{-1/2} \mathbf{\Pi}\widetilde{\mathbf S}_2$, where ${\mathbf S}_2$ and $\widetilde{\mathbf S}_2$ belong to ${\mathcal S}^{K\times 2}$  as described in Definition~\ref{def:udmustm}. If ${\mathbf X}^H_2{\mathbf D}{\mathbf X}_2=\widetilde{\mathbf X}^H_2{\mathbf D}\widetilde{\mathbf X}_2$, then we have ${\mathbf X}_2=\widetilde{\mathbf X}_2$.
	\hfill\QED
\end{proposition}


\subsection{User-constellation Assignment and Power Allocation for the Noncoherent ML Detector}
We consider the user-constellation assignment $\pi$ and power allocation $\mathbf{p}$ for the noncoherent ML detector of design. For our design given in~\eqref{eqn:sigmtx}, we have
\begin{align*}
&\mathbf{X}_2^H \mathbf{D}\mathbf{X}_2 +\sigma^2 \mathbf{I}_2
=\begin{bmatrix}
  \mathbf{s}_1^H \mathbf{s}_1 +\sigma^2 & \mathbf{s}_1^H \mathbf{s}_2 \\
\mathbf{s}_2^H \mathbf{s}_1 & \mathbf{s}_2^H \mathbf{s}_2 +\sigma^2
\end{bmatrix}
=\begin{bmatrix}
a & c\\
c^* & b
\end{bmatrix},\nonumber\\
&\widetilde{\mathbf{X}}_2^H \mathbf{D}\widetilde{\mathbf{X}}_2 +\sigma^2 \mathbf{I}_2
=\begin{bmatrix}
\mathbf{s}_1^H \mathbf{s}_1 +\sigma^2 & \mathbf{s}_1^H \tilde{\mathbf{s}}_2 \\
\tilde{\mathbf{s}}_2^H \mathbf{s}_1 & \tilde{\mathbf{s}}_2^H \tilde{\mathbf{s}}_2 +\sigma^2
\end{bmatrix}=\begin{bmatrix}
a & \tilde{c}\\
\tilde{c}^* & \tilde{b}
\end{bmatrix}.
\end{align*}
where 
\begin{align}\label{eqn:correlationmatrix}
a&=\sum_{k=1}^K \frac{1}{p_k}+\sigma^2,
c=\sum_{k=1}^K s_k, \tilde{c}=\sum_{k=1}^K \tilde{s}_k,\nonumber\\
b&=\tilde{b}=\sum_{k=1}^K p_k |s_k|^2 +\sigma^2=\sum_{k=1}^K E_k d^2 p_k +\sigma^2,
\end{align}
in which $s_k, \tilde s_k \in \mathcal{X}_k$, $\tilde{c}, c\in \mathcal{Q} = \uplus_{k =1}^K \mathcal{X}_k$, and $a b>|c|^2$.
From~\eqref{eqn:correlationmatrix}, we can find that $\mathbf{X}_2^H \mathbf{D}\mathbf{X}_2 +\sigma^2 \mathbf{I}_2$ and $\widetilde{\mathbf{X}}_2^H \mathbf{D}\widetilde{\mathbf{X}}_2 +\sigma^2 \mathbf{I}_2$ are independent of the permutation function $\pi$, but are determined the power allocation vector $\mathbf{p}=[p_1, \ldots, p_K]^T$, and the information carrying vectors $\mathbf{s}=[s_1, \ldots, s_K]^T$ and $\tilde{\mathbf{s}}=[\tilde{s}_1, \ldots, \tilde{s}_K]^T$.
Now, with the help of~\eqref{eqn:correlationmatrix}, we have
\begin{align*}
&\det\Big[(\mathbf{X}_2^H \mathbf{D}\mathbf{X}_2 +\sigma^2 \mathbf{I}_2)(\widetilde{\mathbf X}_2^H \mathbf{D}\widetilde{\mathbf X}_2 +\sigma^2 \mathbf{I}_2)^{-1}) \Big]=\frac{ab-|c|^2}{a \tilde{b}-|\tilde{c}|^2},\\
&{\rm tr}\Big[(\mathbf{X}_2^H \mathbf{D}\mathbf{X}_2 +\sigma^2 \mathbf{I}_2)(\widetilde{\mathbf X}_2^H \mathbf{D}\widetilde{\mathbf X}_2 +\sigma^2 \mathbf{I}_2)^{-1}\Big]\\
&\!=\!\frac{1}{a\tilde{b}-|\tilde{c}|^2}{\rm tr}\Bigg\{\begin{bmatrix}
a & c\\
c^* & b
\end{bmatrix}
\begin{bmatrix}
\tilde{b} & -\tilde{c}\\
-\tilde{c}^* & a
\end{bmatrix}\Bigg\}
\!=\!\frac{ab+a\tilde{b} -c\tilde{c}^*- c^* \tilde{c}}{a\tilde{b}-|\tilde{c}|^2}.
\end{align*}
As a consequence,~\eqref{eqn:KLdistanece} can be reformulated by
\begin{subequations}
	\begin{align*}
	&\mathcal{D}_{\rm KL}(\mathbf{X}_2 ||\widetilde{\mathbf X}_2)= \frac{2ab -c\tilde{c}^*- c^* \tilde{c}}{ab-|\tilde{c}|^2} - \ln\Big (\frac{ab-|c|^2}{ab-|\tilde{c}|^2} \Big )-2\\
	&=\frac{ab-|c|^2}{ab-|\tilde{c}|^2} - \ln\Big (\frac{ab-|c|^2}{ab-|\tilde{c}|^2}\Big )-1+\frac{|c-\tilde{c}|^2}{ab-|\tilde{c}|^2}.
	\end{align*}
\end{subequations}

Recall that, the power constraint in Assumption~\ref{assumption1} is $\mathbb E\{|x_{k,t}|^2\} \le P_k$, $k=1,\ldots, K, t=1,2$. That is, for the first and second time slots, we have
$\mathbb E\{|x_{k,1}|^2\}=\frac{1}{p_{\pi(k)}\beta_{k}}  \le P_k$, and $\mathbb E\{|x_{k,2}|^2\}=\frac{p_{\pi(k)} E_{\pi(k)}d^2}{\beta_k}  \le P_k$. The above power constraints are equivalent to
\begin{align*}
\frac{1}{P_{\pi^{-1}(k)} \beta_{\pi^{-1}(k)}}\le p_k \le \frac{P_{\pi^{-1}(k)} \beta_{\pi^{-1}(k)}}{E_k d^2}, \forall k.
\end{align*}

For the considered massive SIMO when $M$ is large, the PEP will goes to zero exponentially where the exponent determined by the KL distance~\cite{Aazhang03tit}. Since we have one-to-one correspondence between $c$, $\{s_k\}_{k=1}^K$, and $\tilde c$, $\{\tilde s_k\}_{k=1}^K$, we now aim to solve the following optimization problem:
\begin{problem}\label{problem1} Find the optimal power control coefficients $\{p_k\}_{k=1}^K$ and permutation $\pi$, such that:
\begin{subequations}\label{eqn:problem1}
\begin{align}
&\max_{\{p_k\}_{k=1}^K, \pi}\,\min_{\{c, \tilde c\}} ~\mathcal{D}_{\rm KL}(\mathbf{X}_2 ||\widetilde{\mathbf X}_2) \nonumber\\
&\qquad =\underbrace{\frac{ab-|c|^2}{ab-|\tilde{c}|^2} - \ln\Big (\frac{ab-|c|^2}{ab-|\tilde{c}|^2}\Big )-1}_{\rm T_1}+\underbrace{\frac{|c-\tilde{c}|^2}{ab-|\tilde{c}|^2}}_{\rm T_2}\label{eqn:equalpowerdesign1}\\
&{\rm s.t.}~a=\sum_{k=1}^K \frac{1}{p_k}+\sigma^2, b=\sum_{k=1}^K p_k E_k d^2 +\sigma^2,\\ 
&\qquad c=\sum_{k=1}^K s_k,\tilde{c}=\sum_{k=1}^K \tilde{s}_k,\\
&\qquad \frac{1}{P_{\pi^{-1}(k)} \beta_{\pi^{-1}(k)}}\le p_k \le \frac{P_{\pi^{-1}(k)} \beta_{\pi^{-1}(k)}}{E_k d^2}, \forall k.\label{pbm1:powerconstraint}
\end{align}
\end{subequations}
\hfill\QED
\end{problem}


We can observe that~\eqref{eqn:problem1} is a max-min optimization problem where the objective function can be divided into two parts: ${\rm T}_1\ge 0$ and ${\rm T}_2\ge 0$. 

We first consider the inner optimization problem on ${\{c, \tilde c\}}$. It can be observed that the minimum of $\rm T_1=0$ is attained when $\frac{ab-|c|^2}{ab-|\tilde{c}|^2}=1$, or equivalently $|c|=|\tilde{c}|$.  Also, the minimum value of $\rm T_2$ is attained when $c$ and $\tilde{c}$ are the nearest neighboring points and $|\tilde{c}|$ is minimized, e.g., the minimal value of ${\rm T}_2$ can be obtained simultaneously when $c=\frac{(1+j)d}{2}$ and $\tilde{c}=\frac{(1-j)d}{2}$. In this case, we have $|c|=|\tilde{c}|$ and hence ${\rm T}_1=0$ and ${\rm T}_2=\frac{d^2}{ab-d^2/2}=\frac{1}{\frac{ab}{d^2}-\frac{1}{2}}$, where the objective function in~\eqref{eqn:problem1} is a monotonically decreasing against $\frac{ab}{d^2}$. We note that
$\frac{ab}{d^2}= (\sum_{k=1}^K \frac{1}{p_k}+\sigma^2)(\sum_{k=1}^K p_k E_k +\frac{\sigma^2}{d^2})$,
and hence problem~\eqref{eqn:problem1} can be reformulated by:
\begin{subequations}\label{eqn:poweropt1}
	\begin{align}
	&\min_{\{p_k\}_{k=1}^K,\pi}~\Big(\sum_{k=1}^K \frac{1}{p_k}+\sigma^2 \Big) \Big(\sum_{k=1}^K p_k E_k +\frac{\sigma^2}{d^2}\Big)\label{eqn18:objectivefun}\\
	&\quad {\rm s.t.}~
	\frac{1}{P_{\pi^{-1}(k)} \beta_{\pi^{-1}(k)}}\le p_k \le \frac{P_{\pi^{-1}(k)} \beta_{\pi^{-1}(k)}}{E_k d^2}, \forall k.\label{eqn18:const3}
	\end{align}
\end{subequations}

The optimization on Problem~\eqref{eqn:poweropt1} can be carried out by first fixing $\pi$ to find the optimal value of $\mathbf{p}$, and then perform further optimization on $\pi$. To that end, we know from~\eqref{eqn18:const3} that, for any given $\pi$, the feasible range of $d^2$ is given by $d^2 \le \frac{P_{\pi^{-1}(k)} \beta_{\pi^{-1}(k)}}{p_k E_k} \le \frac{P_{\pi^{-1}(k)}^2 \beta_{\pi^{-1}(k)}^2}{E_k}$ for $k=1,\ldots, K$, or equivalently $d^2 \le \min\,\Big\{\frac{P_{\pi^{-1}(k)}^2 \beta_{\pi^{-1}(k)}^2}{E_k}\Big\}_{k=1}^K$.
By the Cauchy-Swartz inequality, we have
\begin{align*}
&\bigg (\sum_{k=1}^K \frac{1}{p_k}+\sigma^2\bigg )\bigg (\sum_{k=1}^K p_k E_k  +\frac{\sigma^2}{d^2}\bigg )\\
&\stackrel{(a)}{\ge} \bigg (\sum_{k=1}^K \frac{1}{\sqrt{p_k}} \sqrt{p_k E_k d^2} +\frac{\sigma^2}{d}\bigg )^2=\bigg (\sum_{k=1}^K  \sqrt{E_k} +\frac{\sigma^2}{d}\bigg )^2,
\end{align*}
where the inequality in~$(a)$ holds if and only if $\frac{\sqrt{p_k E_k}}{{1}/{\sqrt{p_k}}}=\frac{1}{d}$, for $k=1,\ldots K$. Or equivalently, the optimal power allocation is $\mathbf{p}=[p_1^\star, \ldots, p^\star]^T$ where $p_k^\star=\frac{1}{\sqrt{E_k} d}$ for $k=1, \ldots, K$. Our next task is to check the power constraint on $p_k^\star$ given in~\eqref{eqn18:const3} is violated or not. For $d^2 \le \min\,\Big\{\frac{P_{\pi^{-1}(k)}^2 \beta_{\pi^{-1}(k)}^2}{E_k}\Big\}_{k=1}^K$, we have $\frac{1}{p_k^\star}=\sqrt{E_k}d \le P_{\pi^{-1}(k)} \beta_{\pi^{-1}(k)}$, and $p_k^\star E_k d^2=\sqrt{E_k} d \le P_{\pi^{-1}(k)} \beta_{\pi^{-1}(k)}, ~\forall k$,
where no power constraints on are violated for $\mathbf{p}$. Finally, the optimization problem on $\pi$ can be given by
\begin{subequations}
	\begin{align*}
	&\min_{\pi}~\sum_{k=1}^{K} \sqrt{E_k}+\frac{\sigma^2}{d}\quad 
	{\rm s.t.}~d^2\le \frac{P_{\pi^{-1}(k)}^2 \beta_{\pi^{-1}(k)}^2}{E_k}, \forall k.
	\end{align*}
\end{subequations}
Or equivalently, we aim to solve
\begin{align*}
&\max_{\pi}~d
\quad {\rm s.t.}~d^2 \le \frac{P_k^2 \beta_k^2}{E_{\pi(k)}},~\forall k.
\end{align*}
Before proceeding on, we establish the following lemma.
\begin{lemma}\label{lemma:orderedseq}
	Suppose that two positive sequences $\{a_k\}_{k=1}^K$ and $\{b_k\}_{k=1}^K$ are arranged both in a nondecreasing order. If we let	$\Pi$ denote the set containing all the possible permutations of $1,\cdots, K$, then, the solution to
	the optimization problem,
	$\max_{\pi \in \Pi} \min \Big\{ \frac{a_k}{b_{\pi(k)}}\Big\}_{k=1}^K$,
	is given by $\pi^\star(k)=k$ for $k=1,\cdots, K$.~\hfill\QED
\end{lemma}

The proof is omitted due to space constraint. By Lemma~\ref{lemma:orderedseq}, and note that $P_1\beta_1 \le \ldots \le P_k\beta_K$, to maximize $d$, we should let $E_{\pi(1)} \le \ldots \le E_{\pi(K)}$, i.e., the average power of the sub-constellations should be in ascending order. All the above discussions can be summarized into the following theorem:
\begin{theorem}
	The users are ordered such that $P_1\beta_1 \le P_2\beta_2 \le \ldots \le P_k\beta_K$, and we denote $d^\star =\min_{k} \big\{\frac{P_k\beta_k}{\sqrt{E_k}}\big\}_{k=1}^K$, the the optimal power for each user can be given by $\mathbf{p}^\star=[\frac{1}{\sqrt{E_1} d^\star}, \ldots, \frac{1}{\sqrt{E_K} d^\star}]^T$. And the optimal permutation matrix is the identity matrix, i.e., $\mathbf{\Pi}=\mathbf{I}_K$.
	\hfill\QED
\end{theorem}

\section{Simulation Results and Discussion}
In this section, computer simulations are performed to demonstrate the effectiveness of our proposed design in comparison with other existing benchmarks. The small-scale fading is assumed to be the normalized Rayleigh fading. The path-loss $L$ as a function of transmission distance $d$ at antenna far-field can be approximated by
\begin{align*}
10\log_{10} L = 20\log_{10} \Big(\frac{\lambda}{4\pi d_0}\Big ) \!-\! 10\gamma \log_{10} \Big(\frac{d}{d_0}\Big) - \psi, d\ge d_0,
\end{align*}
where $d_0=100$m is the reference distance,  $\lambda=v_c/f_c$ ($f_c=3$GHz) is the wavelength of carrier, $\gamma=3.71$ is the path-loss exponent~\cite{goldsmith05}. In the above model, $\psi \sim \mathcal{N}(0, \sigma_{\psi}^2)$ ($\sigma_{\psi} =3.16$) is the Gaussian random shadowing attenuation resulting from blockage of objects. For the receiver, we assume that the noise power is 
$10\log_{10}\sigma^2 =10\log_{10} N_0 B_w = 10\log_{10} 3.2\times 10^{-10} = -125.97\,{\rm dB}$, where the channel bandwidth $B_w=20$MHz and $N_0= k_0 T_0 10^{F_0/10}$ is the power spectral density of noise with $k_0=1.38\times10^{-23}$ J/K being the Boltzman's constant, reference temperature $T_0 =290$K (``room temperature''), and noise figure $F_0=6$\,dB. 


\begin{figure}[htpb]
	\centering
	\flushleft
	\resizebox{9.5cm}{!}{\includegraphics{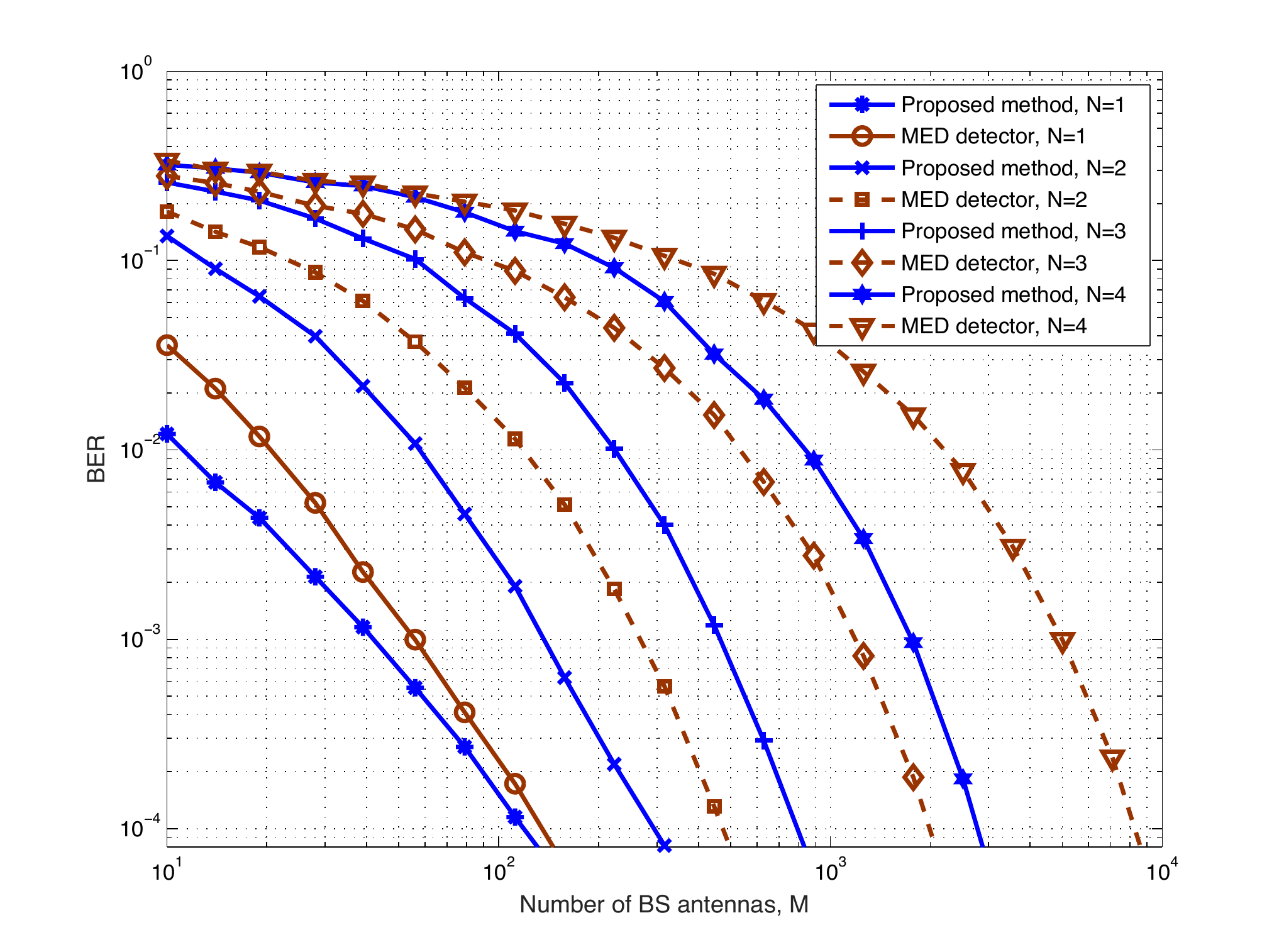}}
	\centering
	\caption{Comparison of the proposed scheme with MED based design on the average BER of all users versus $M$, where 4-QAM are used by all the users.}
	\label{fig:avrbervsdist}
\end{figure}

\begin{figure}[htpb]
	\centering
	\flushleft
	\resizebox{9.5cm}{!}{\includegraphics{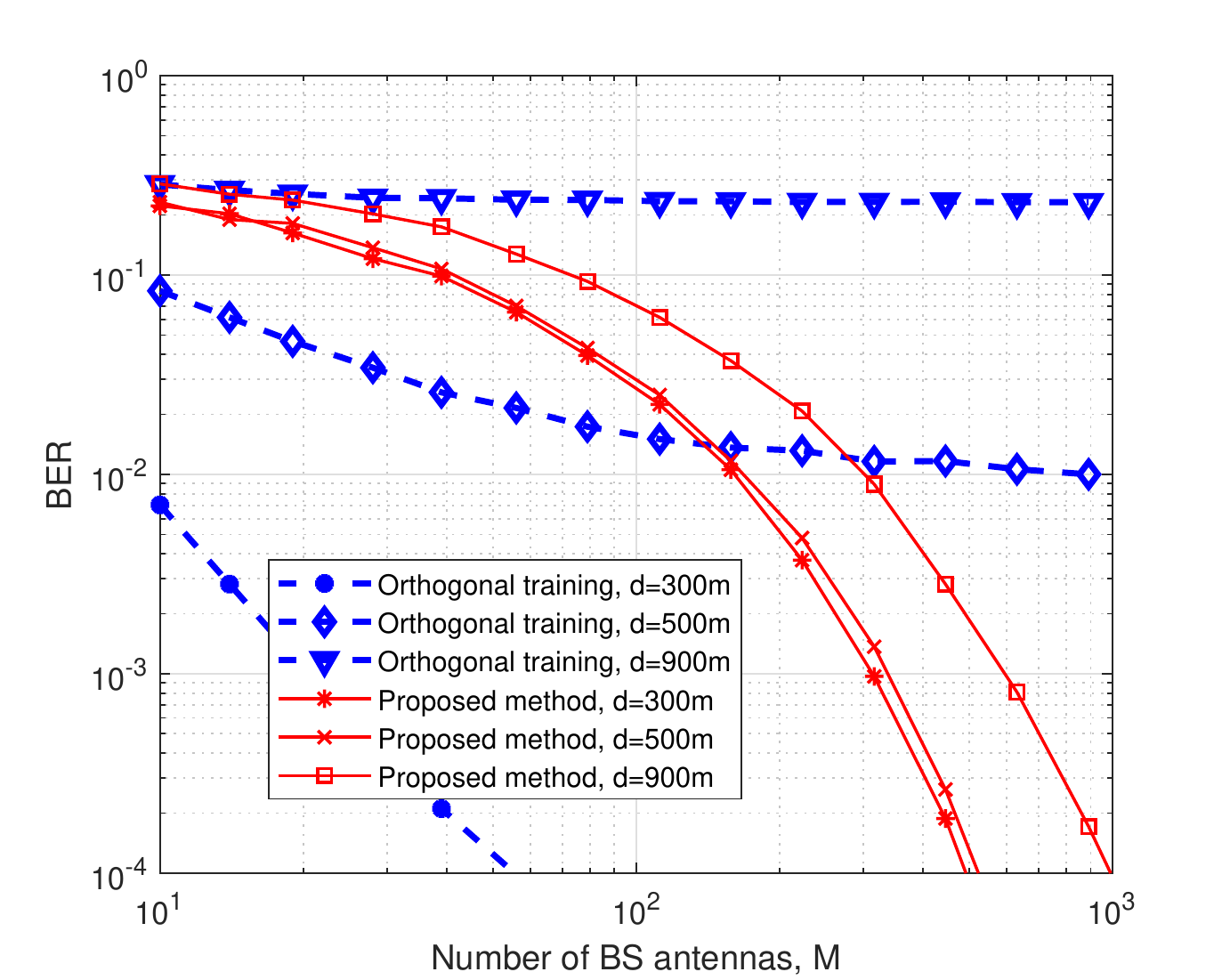}}
	\centering
	\caption{The comparison between the proposed design and ZF receiver with orthogonal channel training for $K=3$ users and 4 time slot.}
	\label{fig:trainingbsproposed}
\end{figure}


We first examine the error performance of the proposed design under the instantaneous average power constraint for different number of users as illustrated in Fig.\,\ref{fig:avrbervsdist}. It is assumed that the average power upper bound is ${P}_k$=316\,mW (25\,dBm), $\forall k$. All the $K$ users are assumed to be uniformly distributed within the cell of radius $d$. It can be observed that, with the increased number of users, the error performance deteriorates quickly caused by the mutual interference between users. Then, more BS antennas are needed to achieve the same average BER.  We also compare our  design with MED based method proposed in~\cite{Goldsmith16tit, zhang2018physically}. Since we are using two time slots, while the MED methods only need one time slot, we assume 2-PAM constellations are used by all the users for the MED based design. We can also find that the proposed approach outperforms the MED based method significantly in terms of BER in all the schemes.

Next, we compare the error performance of the proposed method with the conventional zero-forcing (ZF) receiver using orthogonal training sequence in Fig.\,\ref{fig:trainingbsproposed}. Without loss of generality, we consider a system with $N=3$ users. For the orthogonal training based method, at least 4 time slots are needed and we assume that the channel coefficients are quasi-constant in these times slots. As 4-QAM are used by each user for the proposed scheme, we assume that 64-QAM are assumed for the training based approach in order to achieve a fair comparison. For the training prcocess, we assume that, a popular least-square (LS) channel estimator is employed~\cite{Gershman06}. It can be observed from Fig.\,\ref{fig:trainingbsproposed} that, when the antenna number $M$ is small and the channel gain is large (i.e., the distance $d$ is small), the training based method outperforms the proposed design in term of BER. However, when the antenna number is relatively large, the proposed design has a better error performance, especially at the cell edge.

\section{Conclusion}
In this paper, we have proposed a new noncoherent multiuser massive SIMO design for the low-latency IIoT wireless communication applications. Assuming that the large-scale fading coefficients are known, we presented a simple and systematic construction of the transmitted signal matrix based on the concept of UDCG. In our design, we have used a noncoherent ML receiver over two time slots where no instantaneous CSI is required. To improve the error performance, the minimum KL distance between the received signals corresponding to different transmitted signal matrices was maximized by proper power allocation and sub-constellation assignment. Computer simulations reveal that, our method outperforms significantly the MED based design in terms of average BER. Our approach can also have a better error performance than the orthogonal training design for cell edge users when the array size is large. 
\small
\bibliographystyle{ieeetr}
\bibliography{reference}
\normalsize


\end{document}